\documentclass[9pt,twocolumn,twoside]{osajnl}

\journal{jocn} 
\usepackage{acronym}
\usepackage[normalem]{ulem}
\newcommand{\added}[1]{{\color{black}#1}}

\usepackage{svg}
\usepackage{color}
\usepackage{subcaption}  
\usepackage{float}  
\setboolean{shortarticle}{false}

\acrodef{OLS}{Open Line System}
\acrodef{ROADM}{Reconfigurable Optical Add-Drop Multiplexer}
\acrodef{BVT}{Bandwidth Variable Transceiver}
\acrodef{OSNR}{Optical Signal-to-Noise Ratio}
\acrodef{ASE}{Amplified Spontaneous Emission}
\acrodef{NETCONF}{Network Configuration}
\acrodef{NLI}{Non-Linear Interference}
\acrodef{BER}{Bit Error Rate}
\acrodef{OSaaS}{Optical Spectrum as a Service}
\acrodef{CFO}{Carrier Frequency Offset}
\acrodef{CDC}{Chromatic Dispersion Compensation}
\acrodef{DGD}{Differential Group Delay}
\acrodef{ORP}{Optical Received Power}
\acrodef{SNR}{Signal to Noise Ratio}
\acrodef{PDL}{Polarization Dependent Loss}
\acrodef{ML}{Machine Learning}
\acrodef{ANN}{Artificial Neural Network}
\acrodef{CNN}{Convolutional Neural Network}
\acrodef{OCM}{Optical Channel Monitor}
\acrodef{OPM}{Optical Performance Monitoring}
\acrodef{EDFA}{Erbium Doped Fiber Amplifier}
\acrodef{EON}{Elastic Optical Network}
\acrodef{DWDM}{Dense Wavelength Division Multiplexing}
\acrodef{QoT}{Quality of Transmission}
\acrodef{MLP}{Multi Layer Perception}
\acrodef{SVM}{Support Vector Machines}
\acrodef{AGC}{Automatic Gain Control}
\acrodef{WSS}{Wavelength Selective Switch}
\acrodef{ReLU}{Rectified Linear Unit}
\acrodef{PSD}{Power Spectral Density}
\acrodef{OOK}{On–Off Keying}
\acrodef{XPM}{Cross Phase Modulation}
\acrodef{WDM}{Wavelength Division Multiplexing}
\acrodef{API}{Application Programming Interface}
\acrodef{SLA}{Service Level Agreement}
\acrodef{SDN}{Software Defined Networking}
\acrodef{SOP}{State of Polarization}
\acrodef{OSA}{Optical Spectrum Analyzer}
\acrodef{SELU}{Scaled Exponential Linear Unit}
\acrodef{OLS}{Open Line System}
\acrodef{ROADM}{Reconfigurable Optical Add-Drop Multiplexer}
\acrodef{BVT}{Bandwidth Variable Transceiver}
\acrodef{OSNR}{Optical Signal-to-Noise Ratio}
\acrodef{ASE}{Amplified Spontaneous Emission}
\acrodef{NETCONF}{Network Configuration}
\acrodef{NLI}{Non-Linear Interference}
\acrodef{BER}{Bit Error Rate}
\acrodef{OSaaS}{Optical Spectrum as a Service}
\acrodef{CFO}{Carrier Frequency Offset}
\acrodef{CDC}{Chromatic Dispersion Compensation}
\acrodef{DGD}{Differential Group Delay}
\acrodef{ORP}{Optical Received Power}
\acrodef{SNR}{Signal to Noise Ratio}
\acrodef{PDL}{Polarization Dependent Loss}
\acrodef{ML}{Machine Learning}
\acrodef{ANN}{Artificial Neural Network}
\acrodef{CNN}{Convolutional Neural Network}
\acrodef{OCM}{Optical Channel Monitor}
\acrodef{OPM}{Optical Performance Monitoring}
\acrodef{EDFA}{Erbium Doped Fiber Amplifier}
\acrodef{EON}{Elastic Optical Network}
\acrodef{DWDM}{Dense Wavelength Division Multiplexing}
\acrodef{QoT}{Quality of Transmission}
\acrodef{MLP}{Multi Layer Perception}
\acrodef{SVM}{Support Vector Machines}
\acrodef{AGC}{Automatic Gain Control}
\acrodef{WSS}{Wavelength Selective Switch}
\acrodef{ReLU}{Rectified Linear Unit}
\acrodef{PSD}{Power Spectral Density}
\acrodef{OOK}{On–Off Keying}
\acrodef{XPM}{Cross Phase Modulation}
\acrodef{WDM}{Wavelength Division Multiplexing}
\acrodef{API}{Application Programming Interface}
\acrodef{SLA}{Service Level Agreement}
\acrodef{SDN}{Software Defined Networking}
\acrodef{SOP}{State of Polarization}
\acrodef{OSA}{Optical Spectrum Analyzer}
\acrodef{SELU}{Scaled Exponential Linear Unit}

\title{Interference Detection in Spectrum-Blind Multi-User Optical Spectrum as a Service (Invited ECOC 2024)}

\author[1,*]{Agastya Raj}
\author[2]{Daniel C. Kilper}
\author[1]{Marco Ruffini}

\affil[1]{CONNECT Center, School of Computer Science and Statistics, Trinity College Dublin, Ireland}
\affil[2]{CONNECT Center, School of Engineering, Trinity College Dublin, Ireland}

\affil[*]{rajag@tcd.ie}

\begin{abstract}
With the growing demand for high-bandwidth, low-latency applications, Optical Spectrum as a Service~(OSaaS) is of interest for flexible bandwidth allocation within Elastic Optical Networks~(EONs) and Open Line Systems~(OLS). While OSaaS facilitates transparent connectivity and resource sharing among users, it raises concerns over potential network vulnerabilities due to shared fiber access and inter-channel interference, such as fiber non-linearity and amplifier based crosstalk. These challenges are exacerbated in multi-user environments, complicating the identification and localization of service interferences. To reduce system disruptions and system repair costs, it is beneficial to detect and identify such interferences timely. Addressing these challenges, this paper introduces a Machine Learning~(ML) based architecture for network operators to detect and attribute interferences to specific OSaaS users while blind to the users' internal spectrum details. Our methodology leverages available coarse power measurements and operator channel performance data, bypassing the need for internal user information of wide-band shared spectra. Experimental studies conducted on a 190 km optical line system in the Open Ireland testbed, with three OSaaS users demonstrate the model's capability to accurately classify the source of interferences, achieving a classification accuracy of 90.3\%. 
\end{abstract}
\setboolean{displaycopyright}{false} 

\begin{document}

\maketitle
\footnote{This is a preprint of a paper accepted and published in the Journal of Optical Communications and Networking (JOCN). The final published version is available at: https://doi.org/10.1364/JOCN.551188}
\section{Introduction}
In recent years, there has been a significant rise in data traffic growth due to increasing demands of high-bandwidth, low-latency applications. Optical Networks form the backbone of this shift, providing Tbit/s capacity for data center interconnects. This has led to a shift from fixed-grid networks to \acp{EON} and recently to Open Line Systems, allowing operators to efficiently adjust bandwidth in the network. 

Typically, \ac{DWDM} systems are based on a service model where the operator provides high-capacity optical transmission services by dividing the available fiber bandwidth into multiple channels, each operating at a different wavelength. This has also enabled new flexible transport services like \ac{OSaaS}~\cite{OSaaS_JOCN, OSaaS_ECOC}. \ac{OSaaS} is a developing concept that generally refers to open line systems in which network users are able to operate multiple optical channels (i.e. using their own transceivers), leasing a given spectral window in a fiber link (e.g., 400 GHz), without the overhead of owning and maintaining a full fiber link. Furthermore, this transparent end-to-end connectivity may be realized by using spectrum infrastructure from multiple providers, creating a dis-aggregated networking scenario. This approach has been implemented by an \ac{OSaaS} operator~\cite{geant}, serving only trusted partners, typically other network operators. Expanding \ac{OSaaS} to accommodate a broader user base with enhanced flexibility could unlock new fiber leasing models, leading to more efficient use of excess capacity. Additionally, it enables users to access optical spectrum while bypassing the complexities and costs associated with managing an entire fiber. For instance, mobile operators deploying Open RAN networks in urban areas (e.g., utilizing Split Option 7.2, which demands high-capacity and low-latency links) could benefit from such a model~\cite{wypior_open_2022}.

However, offering shared fiber access to third parties introduces potential security and performance risks. For instance, inter-channel interference, such as crosstalk due to fiber non-linearity~\cite{pointurier_design_2017}, can cause the dynamics of one channel within a given spectral window to negatively impact the performance of others on the same fiber link. This is further exacerbated if the operator uses a dense allocation of spectrum to multiple users, which impacts the channel quality and introduces challenges in identifying which user's signal or activity is the root cause of the interference affecting a given service. Any user-initiated changes such as power adjustments or channel addition, can inadvertently introduce interferences for the operator's own channels, and to other \ac{OSaaS} users. 

Today's operators are cautious of these vulnerabilities in adopting \ac{OSaaS}. In a recent survey with 25 network operators, 60\% of operators were concerned about the power/\ac{PSD} management in \ac{OSaaS}~\cite{kaida_thesis}. Some operators were also cautious of incompatible signals being injected into the network, such as \ac{OOK}. 

A key advantage of \ac{OSaaS} is its ability to allocate user spectrum as a contiguous block, enabling the user to add channels of different bandwidth within that spectral window as required, and simplifying the management overhead for the hosting network operator. Additionally, typically the network operator might not have spectrum analyzers at the ROADMs, thus do not have insight into the internal configuration of the allotted spectrum. In this scenario, the operator would allocate a spectral window on the \ac{ROADM} for the user, and only be able to monitor the overall power across that spectral window from the ROADM's internal \ac{OCM}. The operator however might not have further visibility on the granular \ac{PSD}, number of channels allocated, their modulation schemes, and individual channel power levels. We refer to this as user spectrum blind \ac{OSaaS} and note that channel monitors can be configured to monitor within the user spectral blocks in user spectrum aware configurations. In this work, we focus on the user spectrum blind problem and address spectrum aware considerations elsewhere. 

However, this approach also restricts the operator's ability to detect and address faults or interferences originating from signals within these blocks. This is a challenge for network operators in favor of \ac{OSaaS}, who are seeking accurate and reliable solutions to optimize network performance. In a spectrum-blind configuration, network operators leverage correlations in available data to detect interferences. Given the limited information—such as coarse power data across the bandwidth and end-to-end performance of the operator’s own channels—operators use these as indirect indicators of performance anomalies. As such, developing tools for the prompt detection of interferences caused by \ac{OSaaS} users is critical, even when only partial or incomplete data about the underlying optical links is available.

By targeting the source of interferences, the operator can manage both malicious and non-malicious user behavior, thereby upholding service quality agreements with other users. Although inter-channel interactions are inherently complex, \ac{ML} offers a promising approach for identifying misbehaving channels. Recent studies have applied \ac{ML} to detect failures from the end-user's perspective~\cite{patri_machine_2023} in an \ac{OSaaS} scenario. However, to the best of our knowledge, no study has been done investigating the interferences occurring in \ac{OSaaS} from the operator's perspective. This work specifically focuses on the operator’s challenge of failure detection in the context of \ac{OSaaS} provisioning.

This paper investigates the impact that user-driven changes have on signal interferences in a multi-user \ac{OSaaS} context. We further explore how operators can identify the interference source utilizing only the information accessible to them, eliminating the need for knowledge about the internal user spectrum, which we assume remains undisclosed to the \ac{OSaaS} operator (or would require tapping signals at \acp{ROADM} and deploying \acp{OSA}). We extend our preliminary analysis presented in Ref.~\cite{osaas_ecoc_conf}, by performing experiments in a longer, more representative topology. Additionally, we investigate a crucial issue of potential interference in \ac{OSaaS} networks - introduction of rogue \ac{OOK} channels in networks which are primarily optimized for coherent channels. The main contributions of this work can be summarized as follows:

\begin{figure}[]
  \centering
  \includegraphics[width=0.7\linewidth]{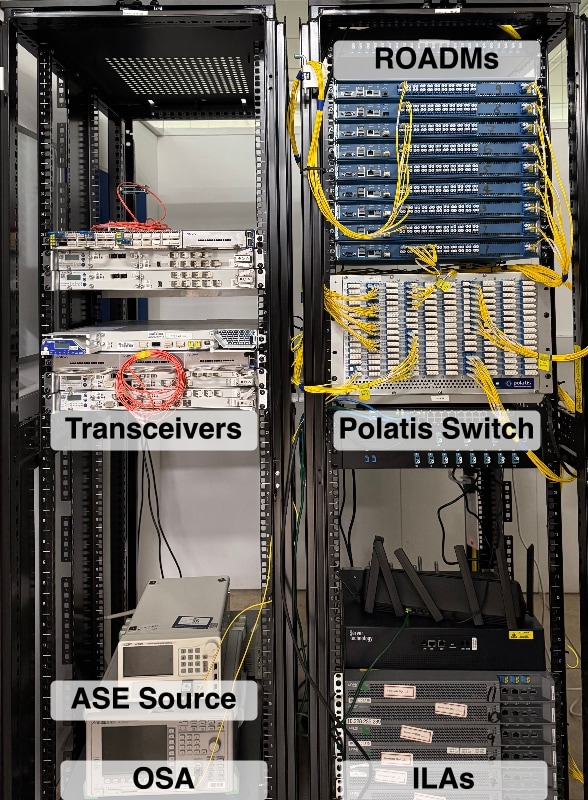}
  \caption{Optical Domain of the Open Ireland Testbed, Dublin, Ireland. Additional fiber spools are connected through the Polatis optical switch.}
\label{testbed}
\end{figure}

\begin{enumerate}
    \item We introduce a framework for operators to use their own data-carrying channels as probes (acting as guard channels) placed between the \ac{OSaaS} windows of different users. 
    \item We examine the interferences that \ac{OSaaS} users can create in the network in a 190 km experimental topology in the Open Ireland testbed. We investigate the effect that power increase, \ac{PSD} limit violations and rogue \ac{OOK} channels can have on the existing users and operator channels. 
    \item We introduce an enhanced 1D-CNN~\added{(One-Dimensional Convolutional Neural Network)} model that leverages data from these operator channels and coarse power measurements at the in-line \acp{ROADM}, to classify the user causing the interference for all three types of impairments. 
\end{enumerate}

The remainder of this paper is organized as follows: Section \ref{chapter_2} describes the experiment setup of the \ac{OSaaS} 
scenario in the Open Ireland Testbed. Section \ref{chapter_3} analyses the effects of different interferences that can be caused in a multi-user \ac{OSaaS} scenario. Section \ref{chapter_4} describes the data collection methodology for the experimental model. Section \ref{chapter_5} describes the proposed 1D-CNN model architecture. Section \ref{chapter_6} discusses the results of the evaluation and finally, Section \ref{chapter_7} summarizes our findings. 

\begin{figure*}[h]
  \centering
  \includegraphics[width=\linewidth]{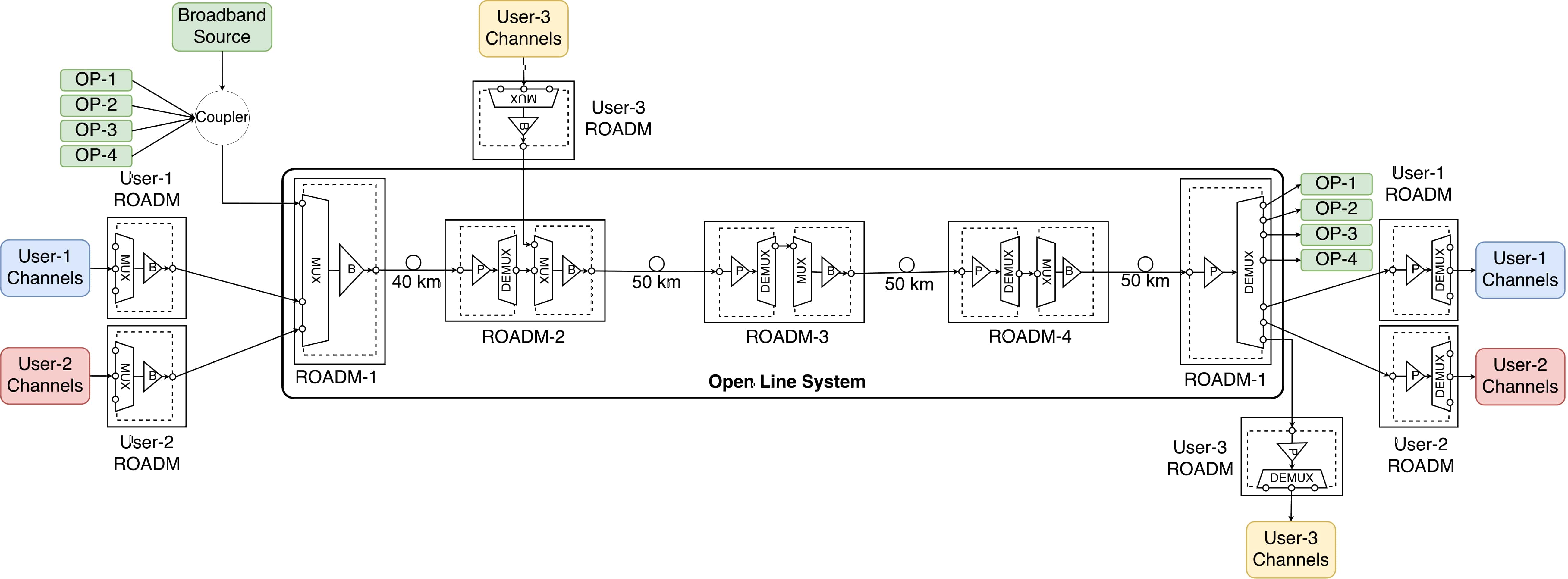}\vspace{-8mm}
\caption{Experimental setup implemented in Open Ireland testbed.}

\label{topology}
\end{figure*}

\section{OSaaS Experimental Setup and Configuration}
\label{chapter_2}

In this section, we describe the \ac{OSaaS} experimental setup and channel configuration using the Open Ireland Testbed.

\begin{figure}[]
\centering
\includegraphics[width=\linewidth]{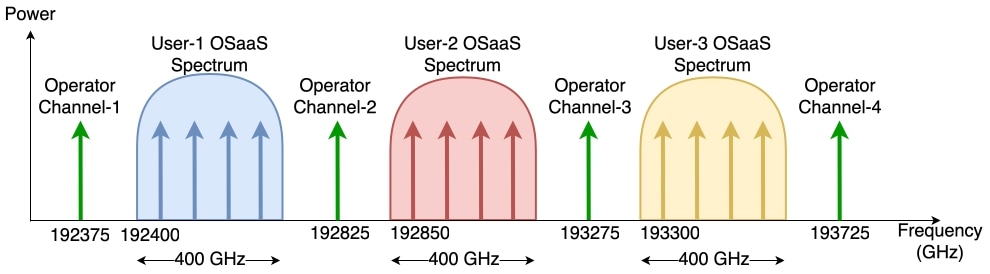}
\vspace{-5mm}
\caption{OSaaS spectrum configuration at steady state for three users.}
\label{spectrum}
\end{figure}

\subsection{Open Ireland Testbed}

The Open Ireland Testbed is a reconfigurable optical wireless programmable testbed in Trinity College Dublin, Ireland, focusing on the interplay between future optical networks and next-generation radio technologies~\cite{open_ireland, EDFA_ECOC}. In particular for the optical domain, the testbed consists of a central Polatis 320 X 320 optical switch, 11 commercial grade Lumentum ROADM-20 units, commercial grade transceivers such as ADTRAN TeraFlex, ADTRAN QuadFlex and Cassini. The testbed also includes over 1700 km of fiber spools, In-Line Amplifiers, broadband sources, \acp{OSA} and standard optical laboratory equipment, some of which are shown in Fig.~\ref{testbed}. The testbed achieves complete reconfigurability through the central Polatis optical switch, which can be configured to dynamically set up different metro-scale topologies for various experiments. All the components can be configured and monitored by a central \ac{SDN} control plane, primarily utilizing the \ac{NETCONF} protocol to dynamically create at-scale topologies, configure network and perform automated data collection. 

\subsection{OSaaS Experimental Setup}

Our experimental setup is implemented in the Open Ireland testbed, depicted in Fig. \ref{topology}. 
The \ac{OLS} consists of four Lumentum \ac{ROADM}-20 units (\ac{ROADM}-1, \ac{ROADM}-2, \ac{ROADM}-3 and \ac{ROADM}-4) configured for uni-directional operation with \ac{WSS} filtering and amplification in each node. The \acp{ROADM} are connected by spans ranging from 40 km - 50 km, with the total length of the topology being 190 km. Four operator channels, acting as the data carrying probe/guard signals, are generated with ADTRAN TeraFlex \acp{BVT}, configured with 400-Gbit/s 64-QAM modulation. The first \ac{ROADM}-1 multiplexes the operator channels, with added \ac{ASE} noise from  a broadband source, in order to collect performance data at different \ac{OSNR} levels of operator probe channels.

For this experiment, we consider the scenario of a sparsely populated spectrum with 3 \ac{OSaaS} end-users. Fig.~\ref{spectrum} shows the \ac{OSaaS} spectrum configuration for the allotted spectra and operator channels. Each user is allocated a spectral window of 400 GHz bandwidth within the C-band in the \ac{OLS}. Since generating coherent signals for each channel for every user would require in excess of 20 transceivers, these user channels were mimicked by shaping an \ac{ASE} broadband source into 50 GHz channels with the edge \acp{ROADM}. However, to gather performance monitoring data for each user, the central frequency of the allotted spectra (\(f_{central}\)) are generated with 3 Cassini transceivers for User-1, User-2 and User-3 respectively, operating at 200 Gbit/s at a modulation format of DP-16QAM. Please note that the monitoring parameters for these user signals are not visible to the operator. 

Each user was allocated an edge \ac{ROADM} to configure channel loading and power levels, and multiplex the different user signals into their allocated OSaaS spectrum. The signal from the operator, along with the user spectra from User-1 and User-2, were multiplexed together in the first \ac{ROADM} of the \ac{OLS}. User-2 spectra is added at the second node, at the ADD port of \ac{ROADM}-2. The operator channels are inserted between different user spectra, and are shown in green color in Fig. \ref{topology}. The signals were equalized at the input \ac{ROADM} MUX, and Booster/Preamp amplifiers were set to a constant gain setting of 18 dB.

\subsection{OSaaS Service Specifications}

\begin{table}[H]
    \centering
    \begin{tabular}{cccc} \hline 
         \textbf{Parameter}&  \textbf{User-1}&  \textbf{User-2}& \textbf{User-3}\\ \hline 
         Start Frequency (GHz)&  192400&  192850& 193300\\ 
         End Frequency (GHz)&  192800&  193250& 193700\\ 
         Optical Bandwidth (GHz)&  400&  400& 400\\ 
         Maximum Power (\(P_{max}\)) (dBm)&  10&  10& 10\\ 
         Maximum PSD (dBm/GHz)&  -15&  -15& -15\\ \hline
    \end{tabular}
    \caption{\ac{OSaaS} requirements sheet with service specifications}
    \label{handover}
\end{table}

\begin{table}[]
    \centering
    \begin{tabular}{ccc} \hline 
         \textbf{Channel}&  \textbf{Frequency}&  \textbf{Modulation}\\ \hline 
         Channel-1&  192375&  DP-64QAM\\ 
         Channel-2&  192825&  DP-64QAM\\ 
         Channel-3&  193275&  DP-64QAM\\ 
         Channel-3&  193725&  DP-64QAM\\ \hline
    \end{tabular}
    \caption{\added{Specifications for operator channels}}
    \label{operator_channel_specs}
\end{table}

Typically, the network operators develop a list of service requirements for the end-user before service handover~\cite{kaida_thesis}. These requirements depend on the network capacity, equipment constraints (such as \ac{EDFA} maximum output power) and the \ac{SLA} with the end-user. 

In this experiment, we develop a standard \ac{OSaaS} handover sheet for all three end users to establish the operational parameters for the service (see Tab.~\ref{handover}). We define the service requirements according to ITU-T Recommendation G.694.1 directive~\cite{itut_694}, assuming standard operation of 200 Gbps 16-QAM channels with a launch power of 0 dBm and channel width of 50 GHz, which is typical for metro networks. The maximum \ac{PSD} and maximum spectral power (\(P_{max}\) are defined at the limit of such channels operating in the 400 GHz spectra, i.e., 8 individual 200 Gbps 16-QAM channels operating at 0 dBm). A margin of $\pm$ 1 dBm is added to account for power fluctuations and other losses.

\begin{figure}[H]
  \centering
  
  \begin{subfigure}[]{\linewidth}
    \centering
    \includegraphics[width=\linewidth]{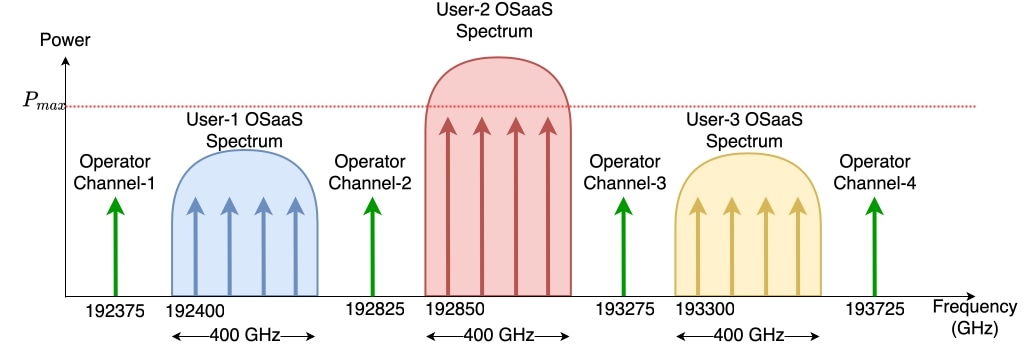}
    \caption{Interference Type 1: Total power increase across the OSaaS User spectrum above the defined \(P_{max}\) in steady state.}
    \label{fig:interference_type_1}
  \end{subfigure}
  
  \vspace{0.5cm}  
  
  \begin{subfigure}[]{\linewidth}
    \centering
    \includegraphics[width=\linewidth]{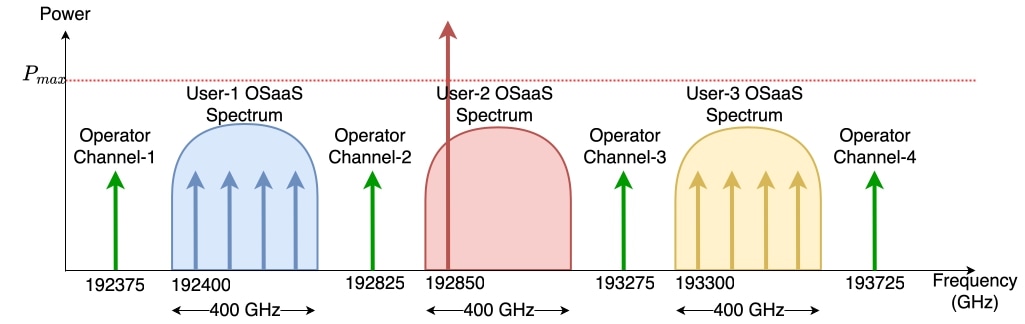}
    \caption{Interference Type 2: The OSaaS user power remains under the \(P_{max}\), but coherent channels are dropped/added, leading to high \ac{PSD}}
    \label{fig:interference_type_2_coherent}
  \end{subfigure}
  
  \vspace{0.5cm} 
  
  \begin{subfigure}[]{\linewidth}
    \centering
    \includegraphics[width=\linewidth]{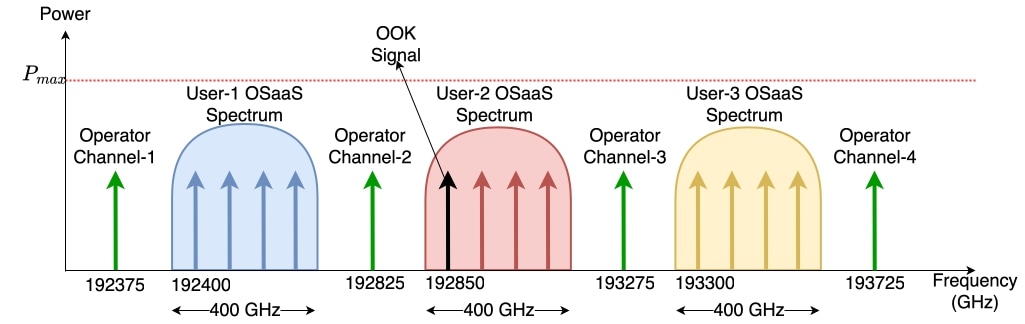}
    \caption{Interference Type 3: The OSaaS user power remains under the \(P_{max}\), but OOK channels are dropped/added.}
    \label{fig:interference_type_3_ook}
  \end{subfigure}
  
  \caption{Different types of interference in the \ac{OSaaS} model.}
  \label{interferences}
\end{figure}

\section{Interferences introduced by OSaaS users}
\label{chapter_3}

In the multi-domain \ac{OSaaS} network environment, the network operator has limited or no control on the way users might load their \ac{OSaaS} spectrum. The only control is typically an attenuation over the entire OSaaS window so that the overall power does not exceed a pre-established threshold. 

To investigate the potential interferences from the \ac{OSaaS} users, we have focused our experiments on three main use cases, as shown in Fig.~\ref{interferences}. We introduced these perturbations for each user into the \ac{OLS}, and measured the end-to-end performance of the operator's channels. We classify a measurement as an interference when there is a drop of more than 0.5 in \(Q\) factor.

\subsection{Increase in power across the \ac{OSaaS} window}

\begin{figure*}[h]
  \centering
    \includegraphics[width=1\linewidth]{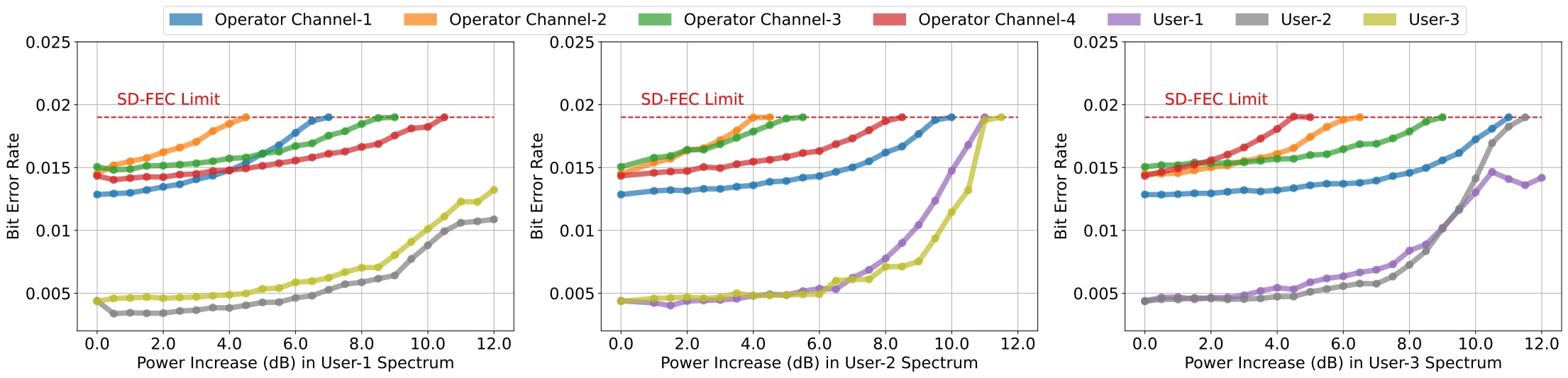}
    \label{pi_interferences_a}
  \caption{Interference Type 1: Total power increase across the OSaaS User spectrum}
\label{pi_interferences}
\end{figure*}

Here we consider the scenario when the user's total spectral power exceeds the Maximum Power $P_{max}$ specified (see Tab.~\ref{handover}). This could happen because of a non-malicious action such as mis-configuration (e.g. on the \ac{EDFA} at the end-user's \ac{ROADM}), or due to malicious actions such as high-power jamming attacks~\cite{jamming}. This can generate interferences due to the \ac{EDFA} cross-talk, caused by the non-flat spectral gain interacting with the \ac{AGC} mechanism, and in part also to the nonlinear crosstalk from nearby channels~\cite{pi_effects}. 
In this case, the user's total spectrum power was systematically increased above the \(P_{max}\) in 0.5 dB increments up to 12 dB, causing a gradual increase in interferences on the other channels. This was achieved by increasing the power of each channel in the \ac{OSaaS} window equally. Fig.~\ref{pi_interferences} shows the impact of increasing total power across each user's allotted spectra, on the nearby users and operator channels. The interfering user causes a significant drop in pre-FEC BER in neighboring channels. This interference is especially visible in the high capacity 400 Gbps 64-QAM operator channels, which go above the SD-FEC threshold after only 5 dB increase in the total power above \(P_{max}\). This can in principle be detected by the operator, by reading the \ac{OCM} value of the entire \ac{OSaaS} window. However we focus our identification only through end-to-end performance monitoring of the operator's channels.

\subsection{ADD/DROP interference}

Although cases of direct increase in overall power across the \ac{OSaaS} window are directly observable, a key issue happens when the overall power is within limit, but the \ac{PSD} is higher because a user might set a smaller number of channels at higher power, increasing the \ac{PSD}. Higher channel power increases inter-channel non-linear interference~\cite{nli}. These interferences are most difficult to identify, as the power levels across the spectral window remain constant, while the performance drops. These cannot be identified by the operator through its \ac{ROADM} \ac{OCM} for the user spectrum blind configuration considered here. 

\begin{figure*}[h]
  \centering
  \includegraphics[width=1\linewidth]{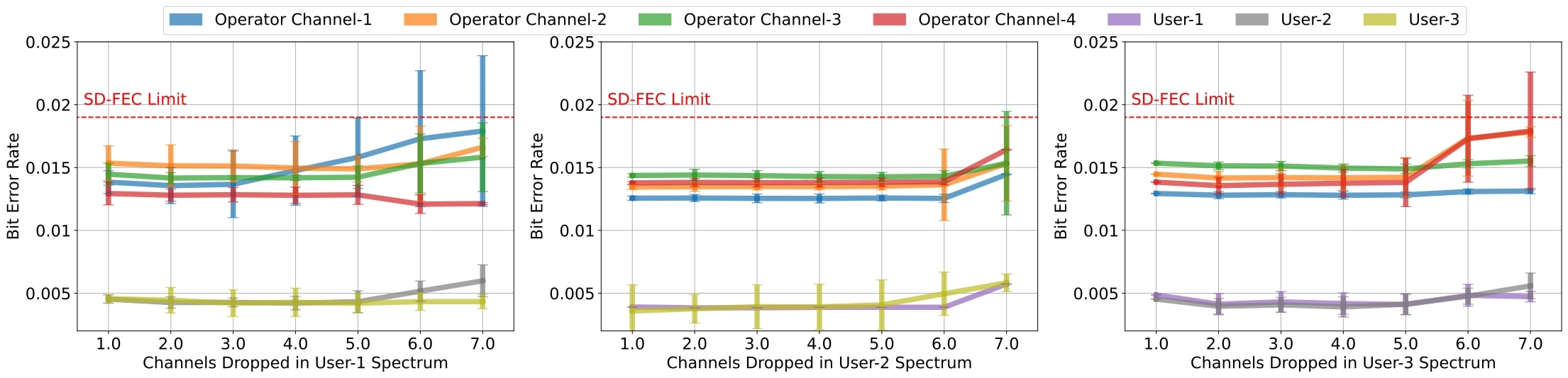}
  \caption{Interference Type 2: The OSaaS user power remains under the \(P_{max}\), but coherent channels are dropped/added. Bars represent the min/max values across all channel configurations.}
\label{ad_interferences}
\end{figure*}

In a fully loaded state, for every 400 GHz spectrum to each user, we configure 8 channels with a channel spacing of 50 GHz, and a channel width of 50 GHz. This is a standard configuration for 16-QAM 200 Gbps signals from commercial transceivers (such as Teraflex or Cassini). Fig.~\ref{ad_interferences} shows the pre-FEC BER of neighboring channels, when each user drops their channels from the initial fully loaded state. It can be observed that there is a significant effect on all channels, especially when a single channel is configured. These interferences can occur due to misconfiguration or a malicious action such as high-power jamming attack. 

\begin{figure*}[h]
  \centering
\includegraphics[width=1\linewidth]{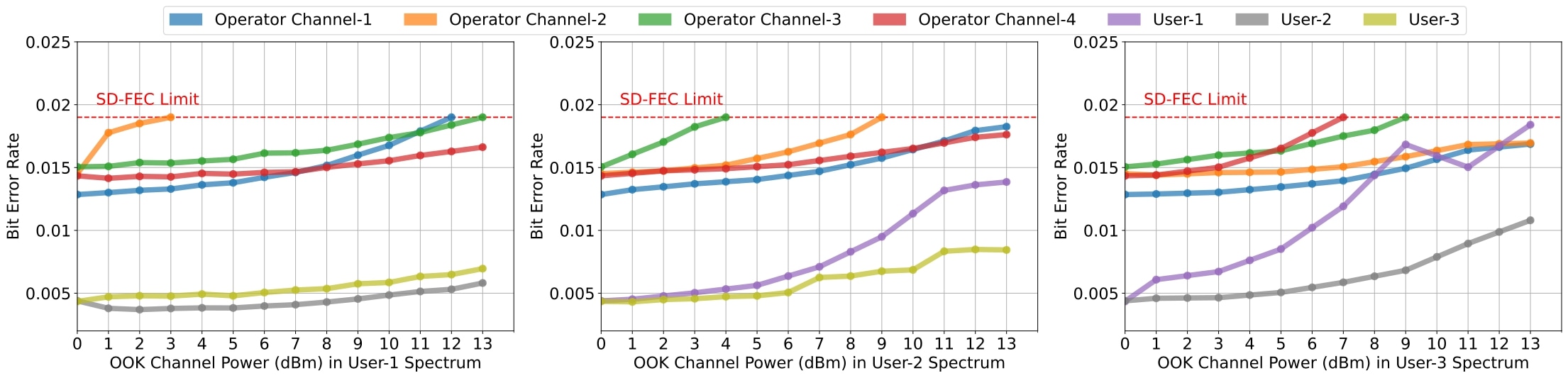}
  \caption{interference Type 3: A single 10 Gbit/s \ac{OOK} channel is inserted into the user's spectrum. The launch power of the \ac{OOK} channel is increased while keeping the total spectrum power below \(P_{max}\).}

\label{ook_interferences}
\end{figure*}
\subsection{Interference due to OOK Channels}

A significant concern for operators interested in implementing \ac{OSaaS} is the possibility of third party customers introducing OOK signals in their coherent systems. The primary limitation in such scenarios stems from the low tolerance of coherent \ac{WDM} channels to \ac{XPM} caused by adjacent intensity-modulated \ac{OOK} channels~\cite{Bononi:09}. Experimental studies have identified \ac{XPM} as the dominant source of interference for coherent channels spaced within 400 GHz of an \ac{OOK} channel. Additionally, a performance degradation is observed even for for channels spaced within 1 THz, attributed to non-linear cross polarization scattering (also known as cross-polarization modulation)~\cite{6204194, Phillips:06}. 
Cross Polarization modulation induces sudden changes in the \ac{SOP} due to intensity modulation of the \ac{OOK} channel. These changes are only weakly dependent on channel spacing and may occur at rates faster than \ac{SOP} tracking in polarization-sensitive transceivers. In coherent transceivers, these rapid \ac{SOP} fluctuations can manifest as amplitude fluctuations, degrading \ac{OSaaS} service performance~\cite{6204194}.

In this experiment, telecoms-grade ADTRAN 10 Gbps \ac{OOK} transceivers were used to introduce \ac{OOK} signals for each user in the \ac{OSaaS} network. At steady state, the launch power of the \ac{OOK} channels was kept at 0 dBm per channel, which is typical of commercial transmission systems using \ac{OOK}. After that, the launch power of the \ac{OOK} channel was increased up to 12 dBm, while maintaining the total user spectrum power below \(P_{max}\). This was achieved by dropping other existing channels in the interfering user’s spectrum, similar to the ADD/DROP interference use case introduced earlier. 

The \ac{OOK} channel was introduced into each user’s allotted spectra, and the effect on existing channels were collected. For each user, we configured the \ac{OOK} channel at 8 different frequencies in the allotted spectrum, and collected the data at different launch powers. Please note that the total power of the user spectra is kept below the threshold of \(P_{max}\), which would make the operator unable to detect misconfigured signals, as the overall power stays the same within the 400 GHz OSaaS window. 

Below, we analyze the effect of these \ac{OOK} channels on the \ac{OSaaS} scenario in our experiment for some specific channel configurations. Fig.~\ref{ook_interferences} shows the impact on pre-FEC BER of neighboring channels (other than the interfering user) for three such scenarios, when the launch power of the \ac{OOK} signal is varied for User-1, User-2 and User-3 respectively. It can be seen that for each user, the neighboring channels are affected the most, due to \ac{XPM} induced by the \ac{OOK} channel. We also observe a significant decrease in performance with higher launch power. This degradation is much higher than changes induced when adding/dropping channels in the previous scenario. In Fig.~\ref{ook_interferences} (left plot), the \ac{OOK} channel is introduced in user-1 spectra at a frequency of 192400 GHz. This causes large effects on the neighboring channels within 400 GHz, such as OP-1 and OP-2; while having non-negligible degradation on the far channels such as user-2, and user-3 channels, as well as OP-3 and OP-4. 
In Fig.~\ref{ook_interferences} (middle plot), the \ac{OOK} channel is introduced in user-2 spectra at a frequency of 193000 GHz. Because of the spectral position, this channel is close to both user spectra, and causes large deterioration at even small launch powers. Similarly, in Fig.~\ref{ook_interferences} (right plot), the \ac{OOK} channel is introduced in user-3 spectra at the starting spectra frequency of 193350 GHz, causing significant deterioration across the spectra. 

\section{Data Collection}
\label{chapter_4}
\begin{table}
    \centering
    \begin{tabular}{ccc}
    \hline
         \textbf{Class}&  \textbf{Total measurements}&\textbf{Event Rate}\\ \hline
         \textbf{No Interference}& 6200  & 85.16\% \\
         \textbf{Power Increase}& 168  & 2.31\%\\
         \textbf{ADD/DROP}& 556  & 7.64\%\\
         \textbf{OOK Channel}&  356 & 4.89\%\\ \hline
    \end{tabular}
    \caption{Measurement statistics for \ac{OSaaS} interference data collection pipeline.}
    \label{tab:my_label}
\end{table}

For each operator channel, we extracted 9 \ac{OPM} features from the Teraflex transceivers, namely \ac{CFO}, pre-FEC BER, \ac{CDC}, \ac{DGD}, optical received power, \ac{OSNR}, Q-factor, \ac{PDL}, \ac{SOP} tracking and the electrical \ac{SNR}. We also collected the power levels of the operator's channels in the \ac{OLS} through in-built \acp{OCM} in the \acp{ROADM}
, and the total output power of each \ac{ROADM}'s ADD, DROP, Line-In and Line-Out Port. It should be noted that in this work, we do not utilize the power readings of the overall \ac{OSaaS} windows for two reasons. Firstly we aim to understand the amount of information an operator can get by simply looking at end-to-end transceiver performance, without using any information from the \ac{OSaaS} channels. Secondly, the channel-related features in the neural network are included as individual channel powers. However, the \ac{OSaaS} window power is an aggregate across all channels, making it unsuitable for our ML model structure. 

\added{Measurements are collected through an automated data collection pipeline, polling all the \acp{ROADM}, transceivers every 30 seconds.} In total, we collect 7280 measurements, out of which there are 1080 cases of interference (i.e., 14.8\% of measurement refer to interference cases with Q-factor drop above 0.5). \added{Since we aim to use only operator-channels related features~(excluding user-related features), and the large amount of combinations of channel configurations, higher number of measurements are needed to cover edge cases}. There are large number of interferences for the ADD/DROP cases, because of large permutations of channel combinations in every user spectrum. 

\section{Model Architecture}
\label{chapter_5}

\begin{figure*}[]
  \centering
  \includegraphics[width=0.75\linewidth]{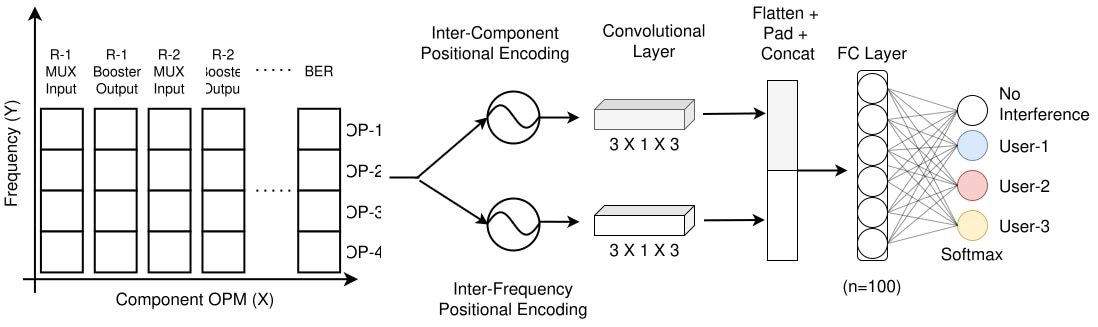}
  \caption{\added{1D-CNN Model Architecture. R-1 and R-2 denote ROADM-1 and ROADM-2, respectively.}}
  \label{model}
\end{figure*}

\added{In a multi-user \ac{OSaaS} environment, timely detection and identification of interference source play a crucial role in upholding service quality. A predictive model can provide a structured method for interference source detection, thereby facilitating mitigation steps by the operator--such as spectrum attenuation and clipping--to maintain optimal network performance. As discussed in the previous section, the complex interaction of optical signals with \ac{ASE} and \ac{NLI} effects demands a model that is not only sensitive to local spectral features but also aware of their absolute positions in the frequency domain. Conventional \ac{ML} approaches for classification, such as Boosted Trees and \acp{ANN}, do not explicitly incorporate positional relationships (e.g. the correlation between spectral positions and signal evolution through the network). To address this limitation, we propose a novel architecture based on a 1D-\ac{CNN} model (Fig. \ref{model}).

Optical spectrum data is inherently sequential because it represents a series of \ac{OPM} measurements taken across the C-band, which in turn are taken along the components through the network. In an \ac{OSaaS} scenario, each operator channel corresponds to a distinct frequency band, and the order of these channels is physically significant for identifying interference sources. Moreover, the order of \ac{OPM} measurements of components is also significant-- since, for example, a new user might be added in the middle of the lightpath (e.g. User-3 in our case). A 1D-CNN can be used to leverage this property through local connectivity and weight sharing. 

For a network with $m$ channels, and $n$ component \acp{OPM}, a single measurement can be viewed as an $(m,n)$ shaped matrix, where the rows correspond to operator channels~(ordered by frequency) and the columns correspond to the ordered \ac{OPM} parameters. This matrix can be interpreted as $m$ parallel 1D sequences~(each of length $n$). For simplicity, let $I(j,k)$ denote the measurement for operator channel index $j$~(with $j\in\{0,...,m-1\}$ and \ac{OPM} index $k$~(with $k\in\{0,...,n-1\}$. Then, a 1D convolution can be applied along the ordered sequence of component \ac{OPM} $k$ , for each fixed $j$ as:
\begin{equation}
    y(j,k) = \sum_{i=0}^{p-1}{w(i)I(j, k+i)} + b,
\end{equation}
where $w(i)$ denotes the convolution kernel of size $p$, and $b$ is the bias term. Here, $y(j,k)$ is a linear combination of local segments $I(j,k), I(j, k+1),...,I(j, k+p-1)$. In contrast, \ac{MLP} treats the input as an unordered vector, by flattening the $(m,n)$ matrix into a vector $\mathbb{x}\in\mathbb{R}^{m.n}$ before applying a fully connected~(FC) layer:
\begin{equation}
z = \sigma\bigl(W, \mathbf{x} + b\bigr), \end{equation} 
where $W\in\mathbb{R}^{H \times(m.n)}$ is the weight matrix for an FC layer with $H$ neurons and $\sigma(.)$ is a non-linear activation function. By flattening, the MLP loses the spatial and sequential arrangement of the data; all $(m.n)$ entries are treated as independent, which makes it harder for the model to learn local patterns (such as interference signatures across neighboring channels). In addition, the 1D-CNN uses the same filter across all positions (weight sharing), thereby reducing the number of parameters and enabling faster training~\cite{ZHANG201937}.

However, standard \acp{CNN} are translation invariant, i.e. models identify given pattern regardless of the absolute position~\cite{cnn_translation_invariance}. In an \ac{OSaaS} interference source detection scenario, the absolute position of an interference event~(e.g. a power spike at 1540 nm versus one at 1550 nm) can be crucial for accurately attributing the source. Similarly, the position along the lightpath~(such as new user being added at ROADM-2) also provide important context. To incorporate absolute positional information, we employ sinusoidal positional encoding~\cite{vaswani_attention_2017} of dimensionality ($d_m=2$) in two parallel branches. 

Essentially, positional encoding adds a deterministic structure that encodes sequence order, adding positional information that convolutional filters alone cannot represent~\cite{JAS-2022-0561}. To capture dependencies across channels as well as component \acp{OPM}, we calculate the feature maps as follows:

\begin{enumerate}
    \item \textbf{Frequency Branch: }For each operator channel index $j$, the positional encoding vector is computed as:

     \begin{equation} 
     PE_m(j, 0) = \sin(j),\quad PE_m(j,1) = \cos(j). 
     \end{equation}
    This encoding can be added to the original measurement~(using broadcasting) to form: 

    \begin{equation}
    I'_m(j,k) = I(j,k) + PE_m(j). 
    \end{equation}

    Here, $I'_m(j,k)$ encodes positional context along the frequency axis. Treating $I'_m(j,k)$ as a 1D sequence~(for each fixed $k$), three distinct 1D convolution layers are applied with a kernel size of 3, resulting in output feature map~($y_m(j,k)$).
    
    \item \textbf{Component Branch: }For each component \ac{OPM} index $k$, the positional encoding vector is computed as:
\begin{equation}
    PE_n(k, 0) = \sin(k),\quad PE_n(k,1) = \cos(k).
\end{equation}
    Similarly, we add this encoding to the measurement: 
    \begin{equation}
        I'_n(j,k) = I(j,k) + PE_n(k)
    \end{equation}
    This encodes the positional context along the component axis. Treating $I'_n(j,k)$ as a 1D sequence~(for each fixed $j$), three distinct 1D convolution layers with a kernel size of 3 are applied, resulting in output feature map~($y_n(j,k)$).
\end{enumerate}

These feature maps from both branches are flattened and concatenated:

\begin{equation}
    y(j,k) = [y_m\left(j,k)\right); y_n\left(j,k)\right)]
\end{equation}
Additionally, a padding of 40 is applied at both ends of the combined feature representation. This padding ensures that, even if additional components or operator channels are added later, the fully connected (FC) layer does not need to be reinitialized. The final feature representation is then fed into a fully connected~(FC) layer with 100 neurons. The \ac{SELU} activation function~\cite{klambauer2017selfnormalizingneuralnetworks} is used after each layer, except for the final softmax output layer, which produces class probabilities over four interference categories. This layer outputs probabilities across 4 classes, aiming to identify if a network state indicates an interference and, if applicable, attributing the interference to the user causing the issue. In this way, the network not only learns local interference patterns through the 1D-CNN’s weight sharing and local connectivity but also retains absolute positional information provided by the positional encodings.
}

The model is trained for 1,200 epochs using the Adam Optimizer, minimizing cross-entropy error. We utilize a batch size of 32 and a decaying learning rate of 0.001. \added{Optimal parameters such as epochs, kernel size, number of neurons and activation functions were identified via a randomized-grid search using stratified K-fold cross validation.} Data splitting for performance evaluation maintained a 3:1 training-to-test ratio, using stratified splits across classes to ensure a balanced class distribution. \added{In total, the training set consists of 5,460 measurements with 810 interferences, and the test set consists of 1,820 measurements with 270 interferences, resulting in an interference rate of 14.8\% across both the datasets.} The models are implemented in PyTorch~\cite{paszke2019pytorchimperativestylehighperformance}, and are trained on an Nvidia GTX 4090 GPU with 24 GB VRAM. 

\section{Results and Discussion}
\label{chapter_6}

\begin{figure}[]
    \centering
    \includegraphics[width=0.53\linewidth]{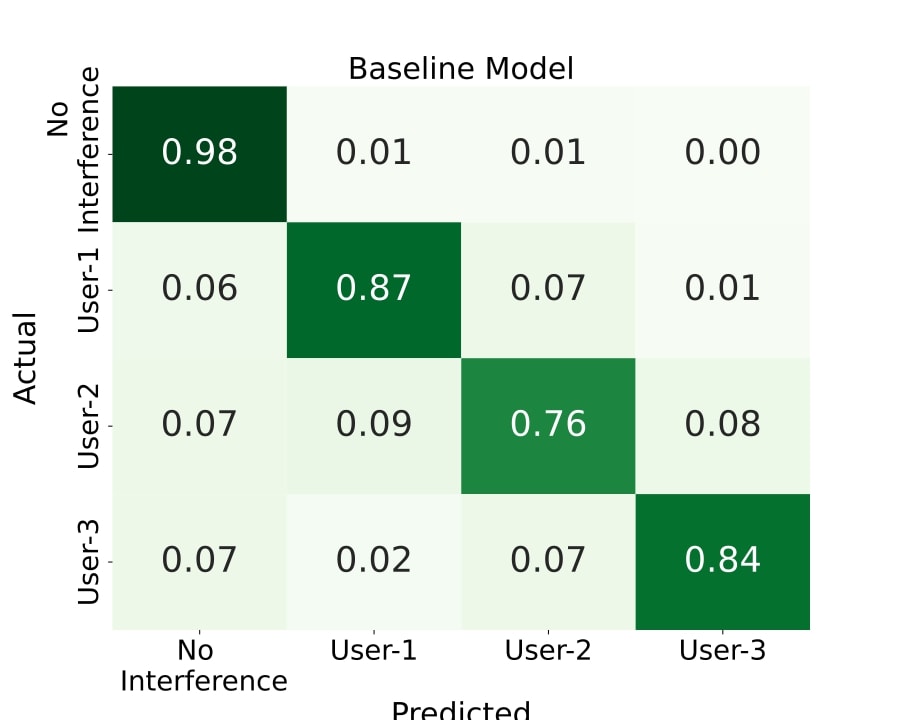}
    \hspace{-7mm}
    \includegraphics[width=0.53\linewidth]{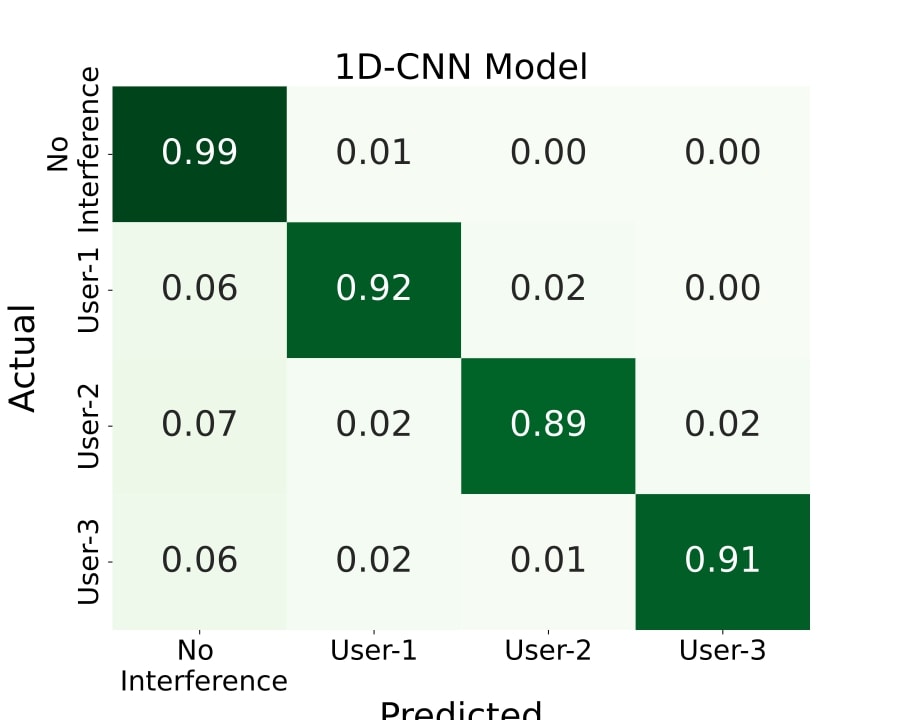}
    \caption{Confusion matrices of interferer classification for: baseline model(left) and 1D-CNN model(right). Values are normalized by the sum of each row.}
    \label{confusion_matrix1}

        \includegraphics[width=
    0.8\linewidth]{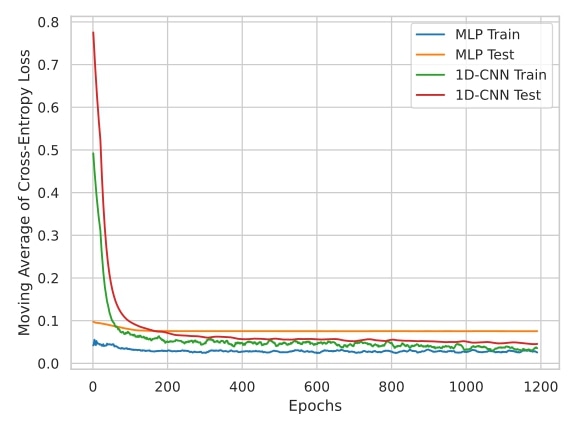}
    \caption{\added{Training Curves for the 1D-CNN and Baseline \ac{MLP} model. Due to large variance in per-epoch training, a moving average of the Cross-Entropy loss is shown.}}
    \label{train_curves}
\end{figure}

\begin{figure*}[]
  \centering
  \includegraphics[width=0.33\linewidth]{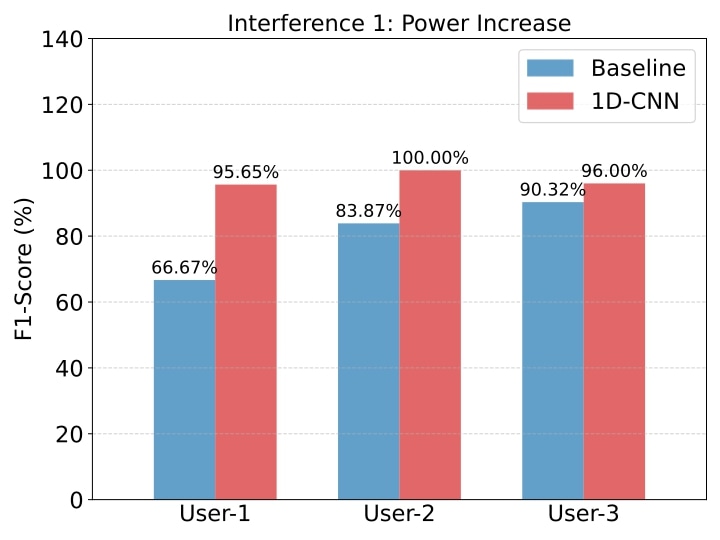}
  \hspace{-3mm}
  \includegraphics[width=0.33\linewidth]{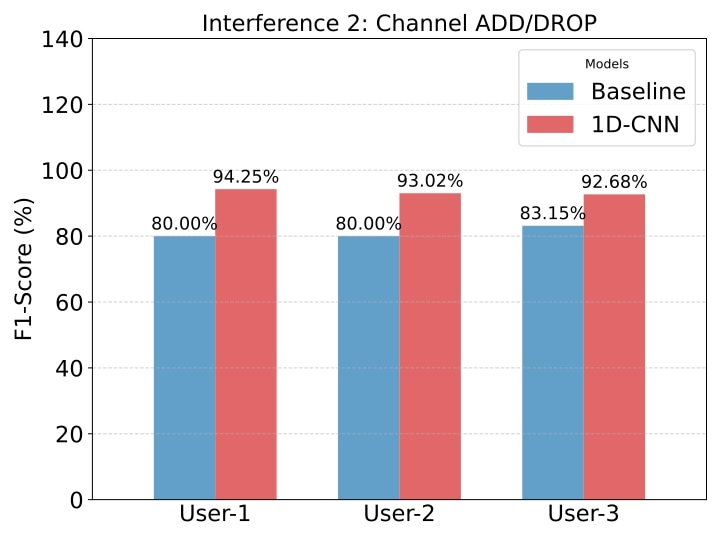}
  \hspace{-3mm}
  \includegraphics[width=0.33\linewidth]{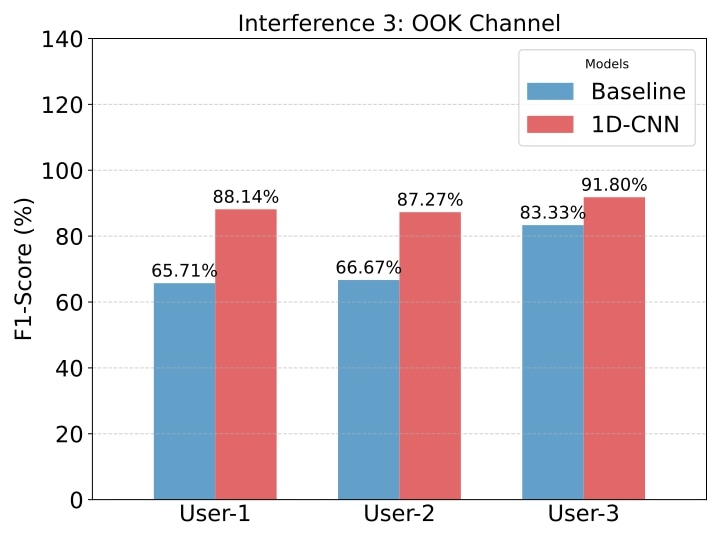}
  \caption{F1-scores for the baseline model and the 1D-CNN model, across User-1, User-2 and User-3, grouped by interference type.}

\label{results}
\end{figure*}

We compare our model with a baseline \ac{MLP} model, using the same set of features to highlight the benefit of our approach. This benchmark model is a 5 layered neural network, with 100 neurons each in the first 4 layers, and 4 neurons in the output layer for classifying interferences across the 4 classes. The first 4 layers of this baseline model employ the \ac{ReLU} activation function, with the output layer using a softmax function to output the probabilities across 4 classes. We also evaluated different benchmark \ac{ML} classification approaches such as \ac{SVM} and Gradient Boosting, however, the \ac{MLP} approach performed the best. A probability threshold of 0.50 is chosen for classification across both the baseline and 1D-CNN model. 

In Fig.~\ref{confusion_matrix1}, the test set's confusion matrix is presented for the baseline \ac{MLP} model and the 1D-CNN model respectively. The 1D-CNN model achieves an overall user classification accuracy of 90.31\%, with a minimum of 89\% for each user. In comparison, the benchmark \ac{MLP} model achieves an overall user classification accuracy of 82.53\%. The performance of the baseline model are below those of our proposed model for all users, and are considerably worse for User-2. It can also be seen that the baseline model has a low accuracy even with a low event rate, especially for User-2, indicating low precision values. 

\added{Figure~\ref{train_curves} shows the training and test losses over 1200 epochs for both the MLP and 1D-CNN models. The MLP converges rapidly to a very low training loss; however, its high test loss indicates significant overfitting. In contrast, while the 1D-CNN model converges more gradually, both its training and test losses decrease steadily, demonstrating better generalization on unseen data.}

Given the interference rate is less than 15\%, precision and recall scores are more relevant metrics than accuracy. A low precision model (with many false positives) would lead to high cost of investigating the issue. However, a low recall model could be more risky because of false negatives, leading to service interruptions. Hence, we evaluate the F1-score for comparing the model's performance. The F1-score is calculated as a combination of both precision and recall:

\[ F1 \space Score = 2 \cdot \frac{Precision \times Recall}{Precision + Recall} \]where
\[Precision = \frac{TP}{TP+FP}, Recall = \frac{TP}{TP+FN}\] where True Positives, False Positives and False Negatives are denoted as TP, FP and FN respectively. Here a positive classification for a respective class denotes the correct classification. 
Fig.~\ref{results} shows the F1-scores across all interference types, for each user. Here, because of multi-class classification, the F1-scores for each user was calculated in a one-vs-all method. For each user, we consider it as positive class and other classes as the negative class. Then for each user, the precision, recall and F1-scores are calculated assuming a binary classification. 

Predictably, the model excels in identifying interference sources causing overall power increase across the \ac{OSaaS} spectrum, due to evident overall power increase across the fiber, although identifying the specific \ac{OSaaS} user generating the impairing is non-trivial. However, even for the second use case, where the total spectrum power remains constant, the model can identify the user causing the interference with high F1-score (between 92\% and 94\% for different users). In comparison, the \ac{MLP} model under-performs in the ADD/DROP type, as those are more difficult to classify. 

The \ac{OOK} channel interference is the most difficult to identify. We attribute this to the fast fluctuations occurring in the tracking parameters because of the intensity modulated \ac{OOK} channel. The 1D-CNN model achieves a minimum of 87\% F1-score across all users, while the baseline model struggles with detecting this impairment. However, the 1D-CNN model consistently outperforms the baseline model by a large margin. This shows the benefit of our model architecture, which takes into account the positions of user and operator spectra, as well as power evolution through the network. 

\added{
Although our experiments were conducted with three active users, the model architecture incorporates several design features that enhance scalability. First, the 1D-CNN kernels act as matched filters by sliding over the input to detect local interference patterns, which makes them reusable across inputs of varying lengths (e.g., when additional components or operator channels are added). Typically, increasing the amount of input data would result in a larger CNN output, potentially requiring re-initialization of the FC layer. To prevent this, padding added after concatenation maintains a fixed input shape for the FC layer, ensuring that even with the addition of more components or users (and thus extra operator channel probes), the FC layer can remain unchanged—as long as the order of components and operator channels is preserved.

Nonetheless, if the number of users increases significantly, it may still be necessary to fine-tune the final output layer with new data to accommodate predictions for the additional users. In practice, \ac{OSaaS} operators (e.g., HeaNET) typically provide a service window for installing and testing new services during which new users can be configured~\cite{OSaaS_JOCN}; this window could be used for additional data collection and model fine-tuning. In future work, we plan to incorporate adaptive learning techniques (such as transfer learning~\cite{EDFA_ECOC}) and evaluate the model in a production network environment with variable user-count scenarios.
}

\section{Conclusion}
\label{chapter_7}

In this work, we examined interference due to power/\ac{PSD} limit violations and rogue \ac{OOK} channel insertion in a multi-user \ac{OSaaS} network. We introduce a framework to enable an operator to identify these malicious or non-malicious behaviors by \ac{OSaaS} end-users, even in a spectrum blind scenario. This is achieved by embedding operator's probe/guard channels between the user spectra to collect monitoring data. We create an experimental metro-scale topology using commercial equipment in the Open Ireland testbed to analyze the interference. Subsequently, we develop a 1D-CNN model to classify the interfering user, using positional encoding to incorporate wavelength and node positions in the network. The results show that the framework achieves a 90.3\% classification rate in predicting the interfering user, with a minimum per user f1-score of 87\%. The framework can accurately locate the source of power/\ac{PSD} violations and rogue \ac{OOK} channels creating interference in a spectrum blind multi-user \ac{OSaaS} scenario. 

\section*{Acknowledgments}
This work was supported by Research Ireland~(RI) (12/RC/2276 p2, 22/FFP-A/10598, 18/RI/5721, ESB Networks through grant 13/RC/2077 p2).

\bibliography{references}

\end{document}